\documentclass[twocolumn, journal]{IEEEtran}
 
\usepackage{booktabs} 
\usepackage{multirow}
\usepackage{mathtools}

\usepackage{circuitikz}

\usepackage{nccmath}

\usepackage{amssymb}
 
\usepackage{enumitem}
\usepackage{stackengine}

\usepackage{smartdiagram}
\usesmartdiagramlibrary{additions}

\graphicspath{{./}}

\usepackage{pdfpages}
\usepackage{pdflscape}
\usepackage{amsmath}
\usepackage{amsthm}

\usepackage{amsfonts}

\usepackage{color}

\usepackage{setspace}
\usepackage{enumitem}
\usepackage{xcolor}

\usepackage{bm}
\usepackage{bbm}

\usepackage{tikz}
\usepackage{pgf}
\usetikzlibrary{arrows,automata}
\usetikzlibrary{shapes}
\tikzstyle{data44}=[rectangle split,rectangle split parts=2,draw,text centered]

\usepackage{subfigure}

\DeclareMathAlphabet{\mathcal}{OMS}{cmsy}{m}{n}

\usetikzlibrary{matrix}

\tikzset{
  BarreStyle/.style =   {opacity=.3,line width=14 mm,color=#1},
  node style ge/.style={},
  node style sp/.style={},
  yl/.style={},
  arrow style mul/.style={},
}

\newtheoremstyle{mytheoremstyle} 
        {\topsep}                    
        {\topsep}                    
        {\fontfamily{ptm}\selectfont}                   
        {}                           
        {\itshape\fontfamily{ptm}\selectfont}                   
        {:}                          
        {.5em}                       
        {}  
\theoremstyle{mytheoremstyle}

\newtheorem{theorem}{Theorem}

\newtheorem{definition}{Definition}

\usepackage{threeparttable,booktabs}

\usepackage[colorlinks,
            linkcolor=blue,
            anchorcolor=blue,
            citecolor=blue
            ]{hyperref}

\usepackage{cite}

\usepackage{mdwlist}

\usepackage[linesnumbered,ruled,vlined]{algorithm2e}
\SetKwInput{KwInput}{Input}                
\SetKwInput{KwOutput}{Output}   
\let\oldnl\nl
\newcommand{\nonl}{\renewcommand{\nl}{\let\nl\oldnl}}

\usepackage{etoolbox}

\makeatletter
\patchcmd{\@algocf@start}
  {-1.5em}
  {0pt}
  {}{}
\makeatother

\usepackage{textcomp}

\usepackage{stmaryrd}

\usepackage{comment}

\usepackage{graphicx,subfigure}

\usepackage{stfloats}
\usepackage{tabularx }

\usepackage{makecell}

\usepackage{colortbl}

\newcolumntype{P}[1]{>{\centering\arraybackslash}p{0.66cm}}
 
\usepackage{anyfontsize}
\usepackage{t1enc}

\newcommand{\T}{^{\scriptscriptstyle\rm T}}

\usepackage[strict]{changepage}
\usepackage{relsize}

\usepackage{diagbox}

\newcommand\tblue{\cellcolor{blue!25}}
\newcommand\tred{\cellcolor{red!25}}

\usepackage{cellspace}

\usepackage{geometry}
\begin{document}


\title{   
    \vspace{-12pt}
    \begingroup
    \fontsize{20pt}{8pt}\selectfont
{A New Neural Network Paradigm for Scalable and Generalizable Stability Analysis of Power Systems}  
\endgroup}

\author{Tong Han,~\IEEEmembership{Member, IEEE}, 
        Yan Xu,~\IEEEmembership{Senior Member, IEEE},
        Rui Zhang,~\IEEEmembership{Member, IEEE}
\vspace{-4pt}
\thanks{Tong Han and Yan Xu (corresponding author) are with the Center for Power Engineering, School of Electrical and Electronic Engineering, Nanyang Technological University, Singapore 639798, Singapore. (e-mail: tong.han, xuyan@ntu.edu.sg)}
\thanks{Rui Zhang is with the School of Electrical Engineering and Telecommunications, UNSW, Sydney, NSW 2052, Australia. (e-mail: rachel.zhang@unsw.edu.au)}

\vspace{-18pt}
}

\maketitle

\begin{abstract} 
    This paper presents a new neural network (NN) paradigm for scalable and generalizable stability analysis of power systems. 
    The paradigm consists of two parts: the neural stability descriptor and the sample-augmented iterative training scheme. 
    The first part, based on system decomposition, constructs the object (such as a stability function or condition) for stability analysis as a \textit{scalable aggregation} of multiple NNs. These NNs remain fixed across varying power system structures and parameters, and are repeatedly shared within each system instance defined by these variations, thereby enabling the \textit{generalization} of the neural stability descriptor \textit{across a class of power systems}. 
    The second part learns the neural stability descriptor by iteratively training the NNs with sample augmentation, guided by the tailored \textit{conservativeness-aware} loss function. The training set is strategically constructed to promote the descriptor's \textit{generalizability}, which is systematically evaluated by verification and validation during the training process.   
    Specifically, the proposed NN paradigm is implemented for large-disturbance stability analysis of the bulk power grid and small-disturbance stability conditions of the microgrid system.  
    Finally, numerical studies for the two implementations demonstrate the applicability and effectiveness of the proposed NN paradigm. 
\end{abstract}


\IEEEpeerreviewmaketitle

\vspace{-10pt}
\section{Introduction} 

Power system stability is a prerequisite for secure and resilient system operation. 
The growing complexity of modern power grids, driven by high penetration of renewable generation, substantial structural variability and expansion of networks, and massive power electronic loads, further challenges the task of system stability analysis \cite{4-rp-8}. 
A power blackout occurred across Spain and Portugal on April 2025 directly triggered by voltage instability, which affected around 60 million people for nearly 10 hours \cite{4-1908}. This recent event again highlights the significance of effective stability analysis methods, serving as the cornerstone for understanding and maintaining system stability.    

The predominant stability analysis methods can be classified into three categories. 
The first category, time-domain simulation-based methods, evaluates stability by performing time-domain simulations under specific disturbances. 
While supporting detailed system models and applicable to various stability problems, it is computationally intensive and lacks general analytical insight \cite{4-1916}. 
The second category, analytical methods, leverages mathematical theories and techniques to extract the fundamental stability characteristics rigorously. For small-disturbance stability, common methods include eigenvalue-based methods \cite{4-614}, impedance-based methods \cite{4-1911}, and nonlinear system theory-based methods \cite{4-1914}. For larger-disturbance stability, the mainstream method is still the Lyapunov method, including its original version \cite{4-1917}, specialized variants (i.e., the energy function (EF) methods) \cite{4-49}, and extensions (e.g., the input-output methods) \cite{4-1919}. 
The third category, learning-based methods, utilizes offline data to learn the mapping from operating conditions to stability conclusions or intermediate representations for stability analysis. This learned mapping is further used for online applications \cite{4-1647}. Since offline learning handles most computational burdens and system models serve only to generate data, this category offers high online efficiency without restricting system models \cite{4-1593}. 
 
Due to the growing scale and dynamic complexity of modern power systems, \textit{scalability} has emerged as a major challenge for stability analysis \cite{4-rp-8}. 
Analytical methods of stability analysis are shifting toward a decentralized pattern for more scalable computation and to facilitate the distributed design of stabilizing controllers. 
Under this paradigm, system stability is verified by examining whether specific conditions are satisfied independently by each group of components within the system. These decentralized stability conditions are analytically derived based on fundamental principles of control theory, including passivity theory \cite{4-rp-12}, dissipativity theory \cite{4-1920}, the small gain theorem and small phase theorem \cite{4-rp-10}, the zero exclusion principle \cite{4-1312}, and the Lyapunov method \cite{4-rp-13}. 
The derivations often require deep insight into system dynamics, which can sometimes be difficult to gain. 
More importantly, scalability is often achieved at the expense of conservativeness, and mitigating this conservativeness remains a critical challenge \cite{4-rp-8}. 
In large-disturbance stability analysis, Lyapunov functions serve as the groundwork for the widely used Lyapunov methods. 
These functions were initially constructed in the form of EFs \cite{4-49}. Due to their analytical formulation and structural consistency across varying systems, EFs are inherently scalable in construction. 
However, their applicability is typically limited to systems satisfying specific assumptions (e.g., Lure-form dynamics and negligible network losses), and they are known to yield overly conservative outcomes in stability analysis \cite{4-1058}. 
To overcome these limitations, various alternative Lyapunov functions have been proposed, such as the polynomial Lyapunov function \cite{4-1741}, the Lyapunov functions family \cite{4-1058}, the adaptive Lyapunov function \cite{4-1917}, and the neural Lyapunov function \cite{4-1926}.  
Their construction generally requires solving optimization problems such as sum of square problems, causing a heavy computation burden for large-scale systems. As a result, scalability is inevitably compromised. 

Regarding learning-based stability analysis methods, \textit{generalizability} is the major challenge they face. 
Unlike analytical methods, which are based on the universal theories or principles applicable to all system operating scenarios, learning-based methods construct stability mappings from data generated by a limited set of offline scenarios \cite{4-1647}. 
However, such data-driven mappings, which essentially capture patterns limited to specific scenarios rather than universal conditions for a class of systems, often struggle to generalize to unseen scenarios, especially those involving network structural changes. 
For instance, the neural Lyapunov functions yielded by learning-based methods may adapt to system parameter variations \cite{4-1862}, but they are invalid if the network topology undergoes any change. 
Some techniques, including model agnostic meta-learning \cite{4-1933}, continuous feature representation \cite{4-1932}, hybrid credibility learning \cite{4-1930}, and graph neural networks \cite{4-1922}, have been introduced to improve the generalizability in learning-based stability analysis methods. However, the achieved generalizability remain limited, as these studies continue to follow a paradigm that does not rely on universal principles and thus inherently lacks the capability to achieve truly generalizable solutions. 

In summary, despite the availability of various stability analysis methods, there remains a gap in scalable and generalizable methods that can readily handle complex system dynamics while maintaining acceptable levels of conservativeness. To this end, we develop a novel neural network (NN) paradigm to enable \textit{scalable and generalizable} stability analysis. 
We refer to the term \textit{stability descriptor} as a general concept covering objects used for stability analysis, including but not limited to stability functions and conditions. The proposed NN paradigm utilizes multiple aggregated NNs to learn stability descriptors via an iterative training scheme. This paradigm offers the following key merits simultaneously: 
\begin{itemize} 
    \item[\textit{1)}]  The paradigm yields stability descriptors that are \textit{generalizable across a class of power systems} with different topological structures, system states and parameters, similar to analytical methods for stability analysis. This strength outperforms the existing learning-based stability analysis methods whose outcomes generally exhibit generalizability only with respect to variations in system parameters or states.
    \item[\textit{2)}]  The paradigm expresses stability descriptors in a \textit{scalable} manner, such that their computation burden remains modest for large-scale power systems. 
    \item[\textit{3)}]  The paradigm has the capability to reduce the \textit{conservativeness} of stability descriptors, a feature that is absent in the existing analytical methods for scalable stability analysis.  
    \item[\textit{4)}]  The paradigm neither relies on strict assumptions about system dynamic models nor requires deep insight into them, similar to the existing learning-based stability analysis methods. This feature is also absent in most analytical methods for scalable stability analysis.
\end{itemize}
These merits are further demonstrated by implementing the proposed NN paradigm for representative problems from both small-disturbance and large-disturbance stability analysis. 

The rest of this paper is organized as follows: Section II presents the proposed NN paradigm; Section III and Section IV implement the NN paradigm in large- and small-disturbance stability analysis, respectively; Section V validates the NN paradigm by numerically testing the two implementations; and Section VI makes a conclusion and a prospect for future works. 

\vspace{-5pt}
\section{Proposed Neural Network Paradigm}

\subsection{Framework}

\vspace{-10pt}
\begin{figure}[h!]
	\centering 
    \includegraphics[scale=0.85]{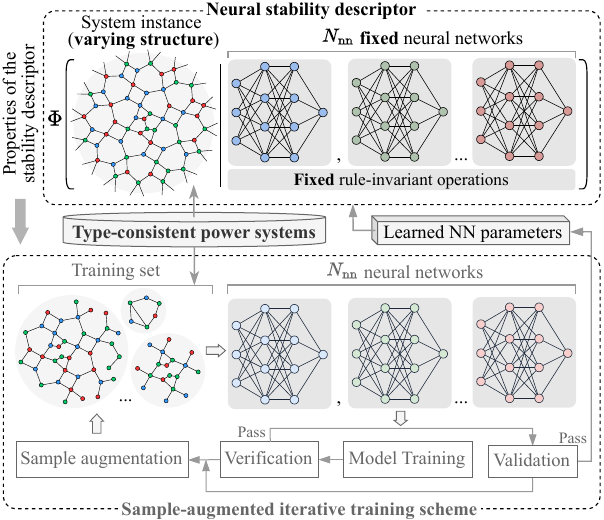}   
    \vspace{-5pt}
	\caption{Framework of the proposed NN paradigm.}
	\label{fig-8-1p2-1} 
\end{figure}

The framework of the proposed new NN paradigm for power system stability analysis is illustrated in Fig. \ref{fig-8-1p2-1}, which consists of the neural stability descriptor and the sample-augmented iterative training scheme. 
Before introducing the framework in details, we first define the concepts of type-consistent power systems and system instances as follows: 

\begin{definition}[\textit{Type-consistent power systems, system instance}]\label{def-8-1p2-1}
    Multiple power systems are considered to be \textit{type-consistent} with one another if they are composed of the same types of buses and branches (i.e., elements that share identical dynamic and steady-state models), even though they may differ in their model parameters and topological structure (i.e., the network size and topology, and the spatial distribution of different types of buses and branches). Within a given class of type-consistent power systems, a \textit{system instance} is uniquely defined by its topological structure, parameters, and states. 
\end{definition}

The neural stability descriptor, denoted by $\Phi(.|.)$, refers to the stability descriptor constructed in the NN paradigm. It is applicable to a class of type-consistent power systems, rather than to a specific system with varying parameters and states. 
The input of the neural stability descriptor $\Phi$ is any system instance from a specific class of type-consistent power systems. 
The descriptor is composed of $N_{\rm nn}$ NNs and rule-invariant operations performed on the descriptor's input and the NNs' outputs. 
Each NN can be shared across different parts within the system. The parameters of these NNs collectively constitute the parameters of the neural stability descriptor. Both the $N_{\rm nn}$ NNs and the rules of rule-invariant operations remain fixed for a specific class of type-consistent power systems. 
The operations are required to be scalable with negligible computational cost as the system size increases. This requirement, combined with the modest computational burden of NNs, ensures that the neural stability descriptor is \textit{scalable}. 

The sample-augmented iterative training scheme is used for learning the parameters of the $N_{\rm nn}$ NNs for a specific class of type-consistent power systems. This scheme consists of five components: the training set, model training, verification, validation, and sample augmentation. 
Each component needs to be designed in accordance with the properties of the specific stability descriptor. 
The \textit{generalizability} of the learned neural stability descriptor is principally attained through three key aspects. The first aspect is that each NN has a low-dimensional input. The second, which is more conventional, involves strategic construction of the training set which consequently yields diverse input instances for each NN. 
The third, and more crucial, factor is the inherent characteristic of the neural stability descriptor, namely that type-consistent systems with varying topological structure and parameters share the same $N_{\rm nn}$ NNs, and each NN is repeatedly shared within a system. 
Moreover, the \textit{conservativeness} of the neural stability descriptor can be optimized by incorporating relevant terms into the loss function $L$.

Accordingly, under this framework, the proposed NN paradigm is capable of addressing a range of power system stability analysis problems, such as constructing stability conditions and finding system stability functions, while ensuring scalability, generalizability, and reduced conservativeness. 
The next sub-sections will elaborate the details of the construction of neural stability descriptors and the sample-augmented iterative training scheme. 

\subsection{Construction of Neural Stability Descriptors}

\begin{figure}[t!]
	\centering 
    \includegraphics[scale=0.88]{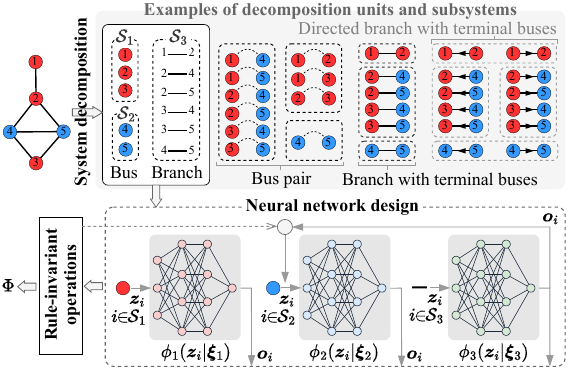}   
	\caption{Schematic diagram of the construction of neural stability descriptors.}
	\label{fig-8-1p2-2}
    \vspace{-5pt}
\end{figure}

The construction of neural stability descriptors for a specific class of type-consistent power systems follows three steps: power system decomposition, NN design, and rule-invariant operations, as illustrate in Fig. \ref{fig-8-1p2-2}. A class of type-consistent power systems composed of two types of buses (indicated in red and blue, respectively) and a single type of branches is used as an example in this figure. 

\subsubsection{Power system decomposition} 
This step first defines the \textit{decomposition units} used for decomposing any system within the given class of type-consistent power systems at the bus and/or branch levels, where multiple decomposition units can be used and the resulting \textit{subsystems} (defined as the components obtained through this decomposition) are allowed to overlap. 
Fig. \ref{fig-8-1p2-2} shows some representative examples of decomposition units and the resulting subsystems. In particular, using directed branches with terminal buses as the unit, each branch is decomposed into two directed branches; using bus pairs as the unit, the system is decomposed into subsystems corresponding to all possible pairs of buses.

Then, based on the defined decomposition units, all possible different subsystems types for the given class of type-consistent power systems are identified. Subsystems with identical dynamic and steady-state models are considered to be of the same type. For instance, for the type-consistent power systems in Fig. \ref{fig-8-1p2-2}, using buses and branches as decomposition units yields three different subsystem types. Using directed branches with terminal buses as the decomposition unit results in four subsystem types, where branch directions may affect sign conventions and thereby the formulations of subsystem models. 

For notation purposes, let $\mathcal{C}$ denote the set of all identified subsystem types, and $c(i) \in \mathcal{C}$ the type of subsystem $i$, and $\mathcal{S}_k$ the set of subsystems of type $k \in \mathcal{C}$. The subsystem of type $k$ is hereafter referred to as the type-$k$ subsystem.   

\subsubsection{Neural network design} 
This step determines the number of NNs needed for constructing the neural stability descriptor, together with the architecture of each NN. 
First, an NN is introduced for each subsystem type $k \in \mathcal{C}$, denoted as:
\begin{equation}
    \bm{o}_i = \phi_k(\bm{z}_i | \bm{\xi}_k )
\end{equation}
where $i \in \mathcal{S}_k$, $\bm{z}_i$ and $\bm{o}_i$ are the input and output vectors of the NN for subsystem $i$, respectively, and $\bm{\xi}_k$ is the combined parameter vector of the NN. 
Thus, the number of NNs equals the number of subsystem types, i.e, $N_{\rm nn} = |\mathcal{C}|$. 

Second, for the architecture of each NN, the input vector to the NN for the type-$k$ subsystems, i.e., $\bm{z}_i$ with $i \in \mathcal{S}_k$, includes at least the properties (including both model parameters and states) of the type-$k$ subsystem which impact the stability descriptor or the associated stability problem. 
The input vector may also include the entries derived from applying certain rule-invariant operations to the NN outputs associated with other subsystems. 
The NN output $\bm{o}_i$ is expected to be scalars, except in cases where a higher-dimensional output is explicitly required. 
Moreover, since each NN is responsible for a subsystem-level static input-output mapping task with low input dimensionality and no spatial dependencies, feedforward neural networks (FNNs) are generally sufficient and appropriate for this setting. 
When an analytical estimation or reference of the NN's output is available, the NN can be designed using a hybrid architecture where an FNN is connected in parallel with the analytical function. 
If the stability descriptor is required to be differentiable, all activation functions must also be differentiable to ensure the overall differentiability of the neural stability descriptor. 

\subsubsection{Rule-invariant operations} 
The rule-invariant operations serve two primary purposes. First, they aggregate the outputs of all NNs, i.e., $\bm{o}_i$ for all $i \in \mathcal{S}_{k}$ and $k\in \mathcal{C}$, or a subset thereof, to produce the final output of the neural stability descriptor. 
Some information from the descriptor's inputs, such as the system size, may also be involved in the aggregation. 
Second, though not always present, they enable the incorporation of outputs $\bm{o}_i$ from certain subsystems into the inputs of others, thereby facilitating information exchange across subsystem.

The operations must satisfy three conditions: (1) the rules of them are fixed for a specific class of type-consistent power systems, ensuring computational uniformity across systems within the class; (2) the operations are invariant to the permutation of inputs corresponding to different subsystems of the same type, given that each NN is expected to generalize across all subsystems of the corresponding type; 
and (3) the operations are scalable, ensuring that the neural stability descriptor remains scalable to varying system sizes. 

\vspace{-5pt}
\subsection{Sample-Augmented Iterative Training Scheme}

Let $\bm{\xi}$ be the vector concatenating the parameter vectors of all $N_{\rm nn}$ NNs, i.e., $\bm{\xi}_k$ for all $k \in \mathcal{C}$. The training scheme, as shown in Fig. \ref{fig-8-1p2-1}, finds the value of $\bm{\xi}$ such that the neural stability descriptor is valid, generalizable, and less conservative. Next, each component of the scheme is elaborated. 

\subsubsection{Training set} 
The training set can be constructed to include sufficient samples from the class of type-consistence systems under consideration, covering a wide range of topological structures, parameter values, and system states. 
This construction yields a sufficient number of diverse input instances for each NN, employed across sufficiently numerous and diverse topological positions within the systems. 
This, in turn, enables the NN to generalize across varying local environments and enhances the generalizability of the neural stability descriptor. 
Moreover, large-scale systems are not necessary for the training set if it contains adequate small- and medium-scale systems.

\begin{figure}[t!]
	\centering 
    \includegraphics[scale=0.86]{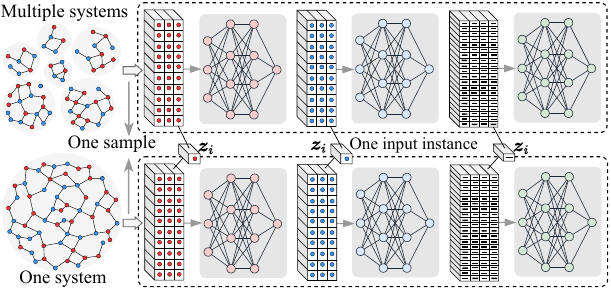}   
	\caption{Illustration of two different construction strategies for the training set.}
	\label{fig-8-1p2-3}
    \vspace{-10pt}
\end{figure}

Alternatively, when the stability descriptor is relatively simple, it may be feasible to obtain a generalizable neural stability descriptor using a training set constructed using only one large-scale system. Fig. \ref{fig-8-1p2-3} illustrates the input instances for each NN, obtained from a sample comprising parameters and system states from multiple small-scale systems, as well as from a larger system. 
The class of type-consistent systems and the system decomposition follow the same setup presented in Fig. \ref{fig-8-1p2-2}. 
It is shown that these two samples result in the same number of input instances for each NN, sourced from different buses or branches. 
Thus, both the aforementioned two strategies for training set construction are capable of generating diverse input instances from a wide range of topological positions within the system. 

\subsubsection{Model Training}
The loss function $L$ is with the following form: 
\vspace{-3pt}
\begin{equation}\label{eq-8-1p2-1}
    \vspace{-3pt}
    L(\bm{\xi}) = w_1 L_1 + w_2 L_2 + w_{\rm aux} L_{\rm aux}
\end{equation}
where $L_1$ directly measures, or serves as a proxy for, the violation of all conditions that the stability descriptor must satisfy; 
$L_2$ measures the conservativeness of the resulting neural stability descriptor; 
$L_{\rm aux}$, though dispensable, offers certain physics-informed insights to the neural stability descriptor, thereby potentially providing guidance during training; $w_1$, $w_2$, and $w_{\rm aux}$ are scalar weights for the loss function components.

Each training round can terminate when the violation measured by $L_1$ is reduced to 0 and the conservativeness measured by $L_2$ is not decreased markedly. 
To facilitate this training objective, the weights can be adjusted strategically within each training round. 
For instance, 
$w_1$ can be adjusted every few training epochs using the Lagrangian duality method \cite{4-1688}; 
$w_2$ can be updated using an exponential growth schedule, with updates occurring each time the violation measured by $L_1$ is reduced to 0 and the epoch number is less than a preset value; and $w_{\rm aux}$ can be updated using an exponential decay schedule. 
In cases where the violation is difficult to reduce to 0, the weight $w_1$ can be set to $1 - w_2$, with $w_2 $ adjusted using an exponential decay schedule. 

\subsubsection{Verification}
The purpose of the verification is to preliminarily verify the generalizability of the neural stability descriptor after training. It may only capture partial generalizability, for instance, with respect to unseen system parameters and states, while excluding unseen topological structures. 
The verification can be designed based on satisfiability problem solving, sufficient sampling, or a hybrid of both, depending on the characteristics of the specific stability descriptor. 
The verification should be able to find counterexamples where the neural stability descriptor is invalid. 
It is noted that if a sample set is used for verification, it should be regenerated for each training round. 

\subsubsection{Validation}
The validation aims to comprehensively examine the generalizability of the neural stability descriptor passing the verification. The validation method can be similar to the verification method; however, it should explicitly consider generalizability to unseen topological structures, and the sample set employed for validation should be substantially larger than that used for the verification. 
This process should consider as many validity indicators of the stability descriptor as possible to ensure a more rigorous validation process. 

\subsubsection{Sample augmentation} 
This process augments the training set by adding counterexamples found by the verification or considering additional topological structures, thereby facilitating generalizability improvement of the neural stability descriptor in the subsequent training round. 
Samples around each counterexample can also be added to the training set to enhancing generalizability to similar violations. 
It is noted that counterexamples found by the validation should not be added to the training set, as this would compromise the integrity of the validation process.

\section{Large-Disturbance Stability Analysis}

This section implements the proposed NN paradigm for large-disturbance stability analysis of the bulk power grid through finding EF-form Lyapunov functions. The stability descriptor in the paradigm is instantiated as the system's EF, and the constructed neural stability descriptor is referred to as the neural EF. 

\subsection{Modeling of System Dynamics and Energy Function}\label{sec-8-1p2-3a}

Consider the class of type-consistent transmission systems that are composed of generator buses with dynamics represented by the classical generator model, load buses with dynamics modeled as a hybrid of constant power and frequency-dependent loads, and transmission lines with power losses ignored. The implementation can also be extended to more complex system dynamics. 
The system consists of $n_{\rm a}$ buses, including $n_{\rm g}$ generator buses and $n_{\rm d}$ load buses, and $n_{\rm b}$ lines. Let $\mathcal{V}$ be bus set, and $\mathcal{E}$ the line set. 
The phase angle of an arbitrary load bus $i^*$ serves as the angle reference. 
All buses are then classified into three types based on their dynamics: type-1 buses including all generator buses, type-2 buses including all load buses except for bus $i^*$, and type-3 buses including bus $i^*$. 
Let $\mathcal{A} \!\!=\!\! \{1, 2, 3\}$ be the set of bus types, and $\mathcal{V}_i$ the set of type-$i$ buses. 
For any bus pair $(i, j)$ with $i, j \!\in\! \mathcal{V}$ and $i \neq j$, its type is defined by the types of the two buses. 
Let $\mathcal{B} \!=\! \{1\text{-}1, 2\text{-}2, 1\text{-}2, 1\text{-}3, 2\text{-}3\}$ collect all bus pair types. 
The dynamics for each type of buses form the state-space system model as follows: 
\vspace{-3pt}
\begin{equation}\label{eq-8-2-1}
    \vspace{-3pt}
    \begin{aligned}
        &\!\!
        \left.   
            \begin{aligned}
                & \dot{\delta_i} \!=\! \omega_i \!-\! \omega_{i^*} \\[-1mm]
                & m_i \dot{\omega}_i \!=\! p_{{\rm g}, i}  \!-\! d_{{\rm g}, i} \omega_i \!-\! b_i \sin(\delta_i \!-\! \theta_i) \\[-1mm]
                & \epsilon \dot{\theta_i} = - b_i \sin(\theta_i - \delta_i) - \textstyle \sum\nolimits_{j \in \mathcal{A}_i}  b_{ij} \sin(\theta_i - \theta_j)
            \end{aligned}
            \right] i \!\in\! \mathcal{V}_1
        \\[-1mm]
        & d_{{\rm d},i} \dot{\theta}_i   \!=\!  \!- p_{\rm d, i} \!-\!  \textstyle {\sum\nolimits_{j \in \mathcal{A}_i}} b_{ij} \sin(\theta_i - \theta_j) \!-\! d_{{\rm d},i} \omega_{i^*} ~~~ i \in \mathcal{V}_2
    \end{aligned}
\end{equation}
where $
    \omega_{i^*} \!\!\doteq\!\! (1 / d_{{\rm d},i^*}) \!\cdot\! (- p_{{\rm d}, i^*} \!-\!\! \textstyle {\sum\nolimits_{j \in \mathcal{A}_{i^*}}} \!\! b_{i^*j} \sin(\theta_{i^*} \!-\! \theta_j))$ and $ \theta_{i^*} \!\doteq\! 0$; 
$\theta_i$ is the voltage angle of bus $i$; 
$\delta_i$ is the rotor angle, 
$\omega_i$ is the rotor angle speed, 
and $p_{{\rm g}, i}$ is the mechanical input power, all of generator $i$; 
$p_{{\rm d}, i}$ is the constant load power, and $d_{{\rm d}, i}$ is the frequency coefficient, both of load $i$; 
$b_i \doteq {e_i u_i}/{x_i}$ with $x_i$ being the transient reactance and $e_i$ the electromotive force of generator $i$, and $u_i$ being the voltage magnitude of bus $i$; 
$b_{ij} \!\doteq\! {u_i u_j}/{x_{ij}}$ with $x_{ij}$ being the reactance of line $ij$; 
$m_i$ and $d_{{\rm g}, i}$ are respectively the inertia and damping coefficients of generator $i$, 
$\mathcal{A}_i$ is the set of buses connecting with bus $i$, 
and $\epsilon$ is a sufficiently small positive numbers introduced by the singular-perturbation approach \cite{4-49}. 
Furthermore, the state-space system dynamic model can be written in a compact form as follows: 
\vspace{-3pt}
\begin{equation}\label{eq-8-2-2}
    \vspace{-3pt}
    \dot{\bm{x}} = f_{s}(\bm{x}; \bm{\rho})
\end{equation}
where $s \!\in\! \mathbb{S}$ with $\mathbb{S}$ being its domain, represents the topological structure of the system; 
$\bm{x} \!\in\! \mathbb{R}^n$ with $n \!=\! 3 n_{\rm g} \!+\! n_{\rm d} \!-\! 1$, is the state vector formed by $\theta_i$ of all buses except for $i^*$, and $\delta_i$ and $\omega_i$ of all generators; 
$\bm{\rho} \!\in\! \mathbb{R}^m\!$ with $m \!=\! 2 n_{\rm d} \!+\! 4n_{\rm g} \!+\! n_{\rm b}$ is the parameter vector formed by $m_i$, $d_{{\rm g}, i}$, $p_{{\rm g}, i}$, $b_i$, $d_{{\rm d}, i}$, $p_{{\rm d}, i}$ and $b_{ij}$ of all associated components; 
and $f_s: \mathbb{R}^{n} \mapsto \mathbb{R}^{n}$ is the vector field of the system with structure $s$. 
The EF of system (\ref{eq-8-2-1}) is defined as follows \cite{4-49}:

\begin{definition}[Energy function]\label{def-8-2-1}
    The EF of system (\ref{eq-8-2-1}) is any differentiable function $V_s\!:\! \mathbb{R}^{n + m} \!\mapsto\! \mathbb{R}$ satisfying the following three conditions: 
    (i)  $\mathcal{L}_{f_s(\bm{x}; \bm{\rho} )} V_s(\bm{x}; \bm{\rho}) \!=\! [\frac{ \partial V_s(\bm{x}; \bm{\rho})  }{ \partial \bm{x} }]\T f_s(\bm{x}; \bm{\rho}) \leq 0$;  
    (ii) if $\bm{x}(t)$ is not an equilibrium point of system (\ref{eq-8-2-1}), the set $\{ t | \mathcal{L}_{f_s(\bm{x}; \bm{\rho} )} V_s(\bm{x}(t); \bm{\rho}) \!=\! 0 \}$ has measure 0 in $\mathbb{R}$; and
    (iii) $V_s(\bm{x}(t); \bm{\rho})$ for any given $\bm{\rho}$ is bounded gives $\bm{x}(t)$ is bounded. 
\end{definition}


\subsection{Construction of the Neural EF}

The existing analytical EFs for system (\ref{eq-8-2-1}) are expressed as the sum of generators' kinetic energy and branches' potential energy. The kinetic energy of a generator depends on its states and parameters, and in some forms (e.g., the EFs in \cite{4-332}), also depends on the states and parameters of the other generators. The potential energy of a branch depends on its parameters and the state variables coupling its two terminals. 
Essentially, the EFs can be viewed as the total energy associated with each bus and each line. 
Accordingly, the decomposition units for constructing the neural EF are defined as buses, branches, and bus pairs. The subsystem type set is then given as follows: 
\vspace{-3pt}
\begin{equation}
    \vspace{-3pt}
    \mathcal{C} = \mathcal{A} \cup \{\rm{e}\} \cup \mathcal{B} = \{1,2,3, \rm{e}, 1\text{-}1, 2\text{-}2, 1\text{-}2, 1\text{-}3, 2\text{-}3 \}
\end{equation}
with type-e subsystems corresponding to line branches. The sets of subsystems of each type are given as follows: 
\vspace{-3pt}
\begin{equation}
    \vspace{-3pt}
    \begin{aligned}
        & \mathcal{S}_1 \!=\! \mathcal{V}_1, \mathcal{S}_2 \!=\! \mathcal{V}_2, \mathcal{S}_3 \!=\! \mathcal{V}_3, \mathcal{S}_{\rm e} \!=\! \mathcal{E}, \mathcal{S}_{1\text{-}2} \!=\! \{(i,j) \!\!\in\!\! \mathcal{V}_1 \!\!\times\!\! \mathcal{V}_2  \}, \\
        & \mathcal{S}_{1\text{-}1} \!=\! \{(i,j) \!\!\in\!\! \mathcal{V}_1 \!\!\times\!\! \mathcal{V}_1  |  i \!\neq\! j  \}, \mathcal{S}_{2\text{-}2} \!=\! \{(i,j) \!\!\in\!\! \mathcal{V}_2 \!\!\times\!\! \mathcal{V}_2  |  i \!\neq\! j  \}, \\
        &  \mathcal{S}_{1\text{-}3} \!=\! \{(i,j) \!\!\in\!\! \mathcal{V}_1 \!\!\times\!\! \mathcal{V}_3   \}, , \mathcal{S}_{2\text{-}3} \!=\! \{(i,j) \!\!\in\!\! \mathcal{V}_2 \!\!\times\!\! \mathcal{V}_3   \},
    \end{aligned}   
\end{equation}

\begin{figure}[t!]
	\centering 
    \includegraphics[scale=0.86]{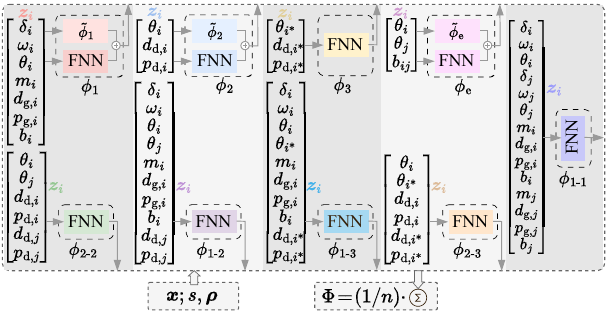}   
	\caption{Illustration of the neural EF.}
	\label{fig-8-1p2-4}
    \vspace{-10pt}
\end{figure}

Following system decomposition, a total of $9$ (i.e., $|\mathcal{C}|$) NNs are introduced to construct the neural EF, including $\phi_{1}(\bm{z}_i|\bm{\xi}_1)$, $\phi_{2}(\bm{z}_i|\bm{\xi}_2)$, $\phi_{3}(\bm{z}_i|\bm{\xi}_3)$, $\phi_{\rm e}(\bm{z}_i|\bm{\xi}_{\rm e})$, $\phi_{1\text{-}1}(\bm{z}_i|\bm{\xi}_{1\text{-}1})$, $\phi_{2\text{-}2}(\bm{z}_i|\bm{\xi}_{2\text{-}2})$, $\phi_{1\text{-}2}(\bm{z}_i|\bm{\xi}_{1\text{-}2})$, $\phi_{1\text{-}3}(\bm{z}_i|\bm{\xi}_{1\text{-}3})$, $\phi_{2\text{-}3}(\bm{z}_i|\bm{\xi}_{2\text{-}3})$, with the architecture illustrated in Fig. \ref{fig-8-1p2-4}. 
The input vector $\bm{z}_i$ for each NN is composed of the entries from the system state vector $\bm{x}$ and $\theta_{i^*}$ that are associated with subsystem $i$, along with the associated entries from the parameter vector $\bm{\rho}$. 
For type-e subsystems, the associated state entries refer to those coupling the two terminal buses, i.e., $\theta_i$ and $\theta_j$ for branch $(i,j)$. 
Each NN produces a single output associated with the energy of its respective subsystem. 

Moreover, analytical EFs for system (\ref{eq-8-2-1}) are available to serve as references for the neural EF. In particular, the following EF taken from \cite{4-49} is adopted as a reference: 
\vspace{-3pt}
\begin{equation}
    \vspace{-3pt}
    \begin{aligned}
        \tilde{V}_s(\bm{x}; \bm{\rho}) =  & \medmath{\sum\nolimits_{i \in \mathcal{V}_1}}  \  [ \tfrac{1}{2} m_{{\rm g}, i} \omega_i^2 - b_i \cos(\delta_i - \theta_i) - p_{{\rm g}, i} \delta_i] \\
         &  - \medmath{\sum\nolimits_{(i,j) \in \mathcal{E}} b_{ij}}   \cos(\theta_i - \theta_j) -  \medmath{\sum\nolimits_{i \in \mathcal{V}_2}}     p_{{\rm d}, i} \theta_i 
    \end{aligned}
\end{equation}
Accordingly, the NNs $\phi_1$, $\phi_2$, and $\phi_{\rm e}$ are constructed as hybrid architectures, each comprising an FNN added to its corresponding analytical reference component: $\tilde{\phi}_1= \frac{1}{2} m_{{\rm g}, i} \omega_i^2 - b_i \cos(\delta_i - \theta_i) - p_{{\rm g}, i} \delta_i$, $\tilde{\phi}_2 = b_{ij} \cos(\theta_i \!-\! \theta_j)$, and $\tilde{\phi}_{\rm e} =  p_{{\rm d}, i} \theta_i$, respectively. 
The remaining NNs are all designed as FNNs. 
Activation functions must be differentiable to ensure differentiability of the neural EF.

Guided by the physical interpretation of EFs, the rule-invariant operation sums the NN outputs of all subsystems and normalizes the result by $\frac{1}{n}$. 
Thus the formulation of the neural EF candidate is constructed as follows: 
\vspace{-3pt}
\begin{equation}\label{eq-8-2-3} 
    \vspace{-3pt}
    V_{s}(\bm{x}; \bm{\rho} | \bm{\xi} ) = \Phi(\bm{x}; s, \bm{\rho} | \bm{\xi}) \!=\!  \frac{1}{n} \medmath{\sum\nolimits_{k\in \mathcal{C}}} \medmath{\sum\nolimits_{i \in \mathcal{S}_k}} \!\! \phi_k(\bm{z}_i | \bm{\xi}_k )  
\end{equation}
where $V_{s}(\bm{x}; \bm{\rho} | \bm{\xi} )$ is a neural realization of $V_s(\bm{x};\bm{\rho})$ in Definition \ref{def-8-2-1}, parameterized by  $\bm{\xi}$.

\subsection{Training of the Neural EF} 

Instead of seeking a global EF, our objective is to construct local EFs defined over specific domains of $\bm{x}$ and $\bm{\rho}$ for all topological structures in $\mathbb{S}$. Local EFs are sufficient if the domains cover all common values of system parameters and post-fault state trajectories of interest. 
Let $\mathbb{X}_s$ and $\mathbb{P}_s$ with $s \!\in\! \mathbb{S}$ be the domains of $\bm{x}$ and $\bm{\rho}$ where the local EFs are valid. 
Then the construction of the neural EF amounts to finding the value of $\bm{\xi}$, such that $V_s$ given by (\ref{eq-8-2-3}) satisfies conditions (i) and (ii) in Definition \ref{def-8-2-1} for all $s \!\in\! \mathbb{S}$, $\bm{x} \!\in\! \mathbb{X}_s$ and $\bm{\rho} \!\in\! \mathbb{P}_s$, and the stability results obtained using $V_s$ are minimally conservative. 
Condition (iii) can be omitted here since we construct local EFs.

\subsubsection{Training set} 
\begin{figure}[t!]
	\centering 
    \includegraphics[scale=0.78]{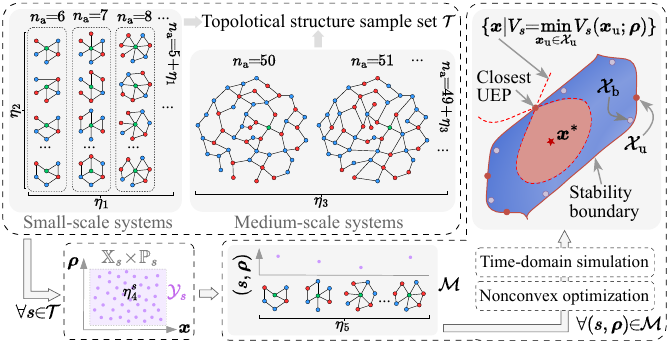}   
	\caption{Illustration of the training set generation.}
	\label{fig-8-1p2-5}
    \vspace{-10pt}
\end{figure}

First, as shown in Fig. \ref{fig-8-1p2-5}, generate a sample set of topological structures from $\mathbb{S}$, denoted as $\mathcal{T}$, which consists of $\eta_1 \times \eta_2$ small-scale topological structures (the system size $n_{\rm a} \!=\! 6, 7, ..., 5 + \eta_1$, and each system size has $\eta_2$ different configurations of network topology and bus types), and $\eta_3$ medium-scale structures ($n_{\rm a} \!=\! 50, 51, ..., 49 + \eta_3$, and each system size has one configuration). 
Second, for each $s \!\in\! \mathcal{T}$, generate $\eta_4^s$ samples of the pair $(\bm{x}, \bm{\rho})$ from the domain $\mathbb{X}_s \!\times\! \mathbb{P}_s$, and collect them into a set denoted by $\mathcal{Y}_s$. 

Third, the training data for the subsequent conservativeness estimation is generated. Specifically, a set $\mathcal{M}$ comprising pairs $(s, \bm{\rho})$ is constructed, where $s$ corresponds to $\eta_5$ topological structure selected from $\mathcal{T}$ with system size $n_{\rm a} \in \{6,7,8\}$. For each selected $s$, one value of $\bm{\rho}$ is drawn from the associated set $\mathcal{Y}_s$. 
For each system with $(s, \bm{\rho}) \in \mathcal{M}$, compute a stable equilibrium point of interest, denoted by $\bm{x}^*$. Then, by solving a series of non-convex optimization problems, identify the set of all type-1 unstable equilibrium points (UEPs), denoted by $\mathcal{X}_{\rm u}$, whose one-dimensional unstable manifold converges to $\bm{x}^*$; 
by time-domain simulations, obtain a set of system states, denoted by $\mathcal{X}_{\rm b}$, which are located inside the region of attraction (RoA) of $\bm{x}^*$ and near its boundary. 
Note that $\bm{x}^*$, $\mathcal{X}_{\rm u}$ and $\mathcal{X}_{\rm b}$ are defined for each pair $(s, \bm{\rho}) \in \mathcal{M}$, but their superscripts are omitted for notational simplicity.

\subsubsection{Model Training}

According to Eq. (\ref{eq-8-1p2-1}) in the proposed NN paradigm and conditions (i) and (ii) in Definition \ref{def-8-2-1}, the loss function $L$ for training all NNs of the neural EF is defined as follows: 
\vspace{-3pt}
\begin{equation}\label{eq-8-2-4} 
    \vspace{-5pt}
    L(\bm{\xi}) \!=\! (1 \!-\! w_2) \underbrace{ \frac{1}{ |\mathcal{T}|} \! \medmath{\sum\limits_{s \in \mathcal{T}}} \frac{1}{|\mathcal{Y}_s|} L_1^s }_{ \eqqcolon L_1 } + w_2 \underbrace{ \frac{1}{|\mathcal{M}|} \medmath{\sum\limits_{(s, \bm{\rho}) \in \mathcal{M}} } L_2^{s, \bm{\rho}}  }_{  \eqqcolon L_2  }   
\end{equation}
with 
\vspace{-3pt}
\begin{equation}\label{eq-8-2-5} 
    \vspace{-3pt}
    L_1^s \!=\!\!\!\!\!\!\! \medmath{\sum_{(\bm{x}, \bm{\rho}) \in \mathcal{Y}_s }} \!\!\!\!\!\! \max( 0, \mathcal{L}_{f_s} V_s ) \!+\! h_{\varphi}(\!  \frac{\Vert f_s \Vert_2}{n}  \!) \cdot \max(0,\!  \frac{\mathcal{L}_{f_s}\! V_s}{n}  + \varphi'  )
\end{equation}
\begin{equation}\label{eq-8-2-6} 
    \vspace{-3pt}
    L_2^{s, \bm{\rho}} =  \frac{1}{ |\mathcal{X}_{\rm b}| } \medmath{\sum\nolimits_{\bm{x} \in \mathcal{X}_{\rm b} }} \Big[ \frac{V_s - \min\nolimits_{\bm{x}_{\rm u} \in \mathcal{X}_{\rm u}} V_{s}(\bm{x}_{\rm u}; \bm{\rho} | \bm{\xi} )}{ V_s - V_s( \bm{x}^*; \bm{\rho} | \bm{\xi} ) } \Big]^2
\end{equation}
where $\varphi$ and $\varphi'$ are small positive numbers; 
$h_{\varphi}\!: \!\mathbb{R} \!\mapsto\! \mathbb{R}$ is define as 0 when the input is less than $\varphi$, and as 1 otherwise; 
we write $f_s$ and $V_s$ instead of $f_s(\bm{x}; \bm{\rho})$ and $V_s(\bm{x}; \bm{\rho} | \bm{\xi})$ for notational simplicity. 
The term $L_1^s \!\geq\! 0$ measures the violation of condition (i) and (ii) for topological structure $s$ over the sample set $\mathcal{Y}_s$; and when $L_1=0$, the neural EF satisfies the two conditions for all topological structures, system parameters, and states samples in the training set. 
The term $L_2^{k, \bm{\rho}}$ measures, for structure $s$ with parameter $\bm{\rho}$, the average relative deviation of the neural EF values at the closest UEP and points near the RoA boundary; and smaller $L_2$ values indicate that the neural EF provides RoA estimates closer to the true RoAs, implying lower conservativeness. 
In addition, the weight $w_2$ is adjusted following an exponential decay schedule. 
Each training round terminates when $L_1$ equals 0 for two consecutive epochs.

\subsubsection{Verification} 
The verification of the neural EF with the trained NNs consists of two parts: formal verification and empirical verification. 
Formal verification verifies the validity of the obtained neural EF over the entire subdomains of $\bm{x}$ and $\bm{\rho}$ for each small-scale topological structure in set $\mathcal{T}$. Specifically, it solves the following satisfiability problem: 
\vspace{-3pt}
\begin{equation}\label{eq-8-2-7} 
    \vspace{-3pt}
    \begin{bmatrix}
        \bm{x} \!\in\! \mathbb{X}_s \land \\ 
        \bm{\rho} \!\in\! \mathbb{P}_s
    \end{bmatrix}
    \land
    \begin{bmatrix}
        ( \mathcal{L}_{f_s} V_s \!\geq\! \varepsilon  \land  \Vert f_s \Vert_2 \!\leq\! n \cdot \varphi  ) \lor \\
        ( \mathcal{L}_{f_s} V_s \!\geq\! 0 \land \Vert f_s \Vert_2 \!\geq\! n \cdot \varphi )
    \end{bmatrix}
\end{equation}
with $\varepsilon > 0$ being the error tolerance. Solving (\ref{eq-8-2-7}) can identify counterexamples of $\bm{x}$ and $\bm{\rho}$ where condition (i) or (ii) fails for the neural EF, or encounter infeasibility, thereby verifying the validity of the neural EF for topological structure $s$. 

While formal verification offers rigorous validity guarantees, its computational burden increases significantly with increasing system size. 
Accordingly, empirical verification is performed for each medium-scale topological structure in set $\mathcal{T}$. 
It evaluates the satisfaction of condition (i) or (ii) utilizing sufficient samples from $\mathbb{X}_{s} \times \mathbb{P}_s$, to identity counterexamples or verify the validity of the neural EF for topological structure $s$. 
Given the inherent characteristics (i.e., rule-invariant operations and fixed NNs across all type-consistent systems) of the neural EF, if it has rigorous and empirical validity guarantees for a sufficient number of small-scale and medium-scale systems, respectively, the rigorous validity guarantee for the medium-scale systems can be almost assured. 
The verification is passed if the validity of the neural EF is verified for all topological structures in the set $\mathcal{T}$.

\subsubsection{Validation} 

The validation set is first constructed, comprising $\eta_6$ small-scale topological structures and $\eta_7$ large-scale topological structures ($n_{\rm a} \geq 200$), none of which are included in the training set. For each large-scale structure, $\eta_8$ samples of $(\bm{x}, \bm{\rho})$ are generated from $\mathbb{X}_{s} \times \mathbb{P}_s$. 
The validity of the obtained neural EF w.r.t. an arbitrary topological structure is assessed similarly to the verification but using the validation set. 
The validation is passed if the neural EF is valid for all topological structures in the validation set.  

\subsubsection{Sample augmentation} 
Following the failed verification, sample augmentation adds the identified counterexamples of $\bm{x}$ and $\bm{\rho}$, along with $\eta_9$ neighboring samples for each to the associated set $\mathcal{Y}_s$ of the training set. 
This augmentation helps improve the validity of the neural EF for all topological structures in set $\mathcal{T}$ of the current training set.  
Following the failed validation, new topological structures and their associated sample sets $\mathcal{Y}_s$ are added to the training set. 
Specifically, let $n_{\rm as}$ and $n_{\rm am}$ be the largest system size of small-scale and medium-scale topological structures in the current training set, respectively. 
The added topological structures consist of $\eta_2$ small-scale structures, each with an identical system size of $n_{\rm as}+1$ but different configurations of network topology and bus types; and $\eta_{10}$ medium-scale structures with different system sizes ranging from $n_{\rm am} + 1$ to $n_{\rm am} + \eta_{\rm 10}$.

\section{Small-Disturbance Stability Analysis}\label{sec-8-1p2-4}

This section implements the proposed NN paradigm for small-disturbance stability analysis of the microgrid system.  
The stability descriptor in the paradigm is instantiated as the decentralized stability condition (DSC). The stability condition derived through the NN paradigm is not only computationally scalable but also structurally decentralized. 
The constructed neural stability descriptor is referred to as the neural DSC. 

\subsection{Small-Disturbance Model and Stability Conditions}

Consider the class of type-consistent microgrids which are composed of generator buses connected with inverters with $p$-$\omega$/$q$-$v$ droop control, load buses connected with $\omega$- and $v$-dependent loads, and low-voltage lines. 
The implementation is extendable to more complex inverter and load dynamics. 
We adopt the same notation as in Section \ref{sec-8-1p2-3a}, including $n_{\rm a}$, $n_{\rm g}$, $n_{\rm d}$, $n_{\rm b}$, $\mathcal{V}$, $\mathcal{E}$, $\mathcal{V}_i$, and $\mathcal{A}$. 
It is noted that $\mathcal{A} \!\!=\!\! \{1,2\}$ for this implementation since we classify all buses into type-1 buses including all generator buses and type-2 buses including all load buses. Moreover, let $\mathcal{D} \!\!=\!\! \{(i, j), (j, i) | (i, j) \!\in\! \mathcal{E} \}$ be the set of all possible directed branches. 
For each directed branch $(i, j) \!\in\! \mathcal{D}$, its type is determined by the types of two terminal buses and branch direction. 
Define the set of all possible directed branch types as $\mathcal{F} \!\!=\!\! \{1\text{-}1, 2\text{-}2, 1\text{-}2, 2\text{-}1\}$, where, for example, $1\text{-}2$ denotes a branch directed from a type-1 bus to a type-2 bus. 
The small-disturbance dynamic model of the microgrid can be formulated as the collection of individual dynamics at all buses $i \in \mathcal{V}$ as follows:
\begin{equation}\label{eq-8-1-2} 
        \bm{E}_{c(i)} \Delta \dot{\bm{x}_i}  \!=\!\!  \bm{A}_{c(i)}(  \bm{\rho}_i ) \cdot \Delta  \bm{x}_i +\! \medmath{\sum\nolimits_{j \in \mathcal{A}_i}} \!\! \bm{A}_{c(i) c(j)} (\bm{\rho}_{i}, \bm{\rho}_{ij}) \! \begin{bmatrix}
            \Delta  \bm{x}_i \\[-1mm] \Delta \bm{x}_j
        \end{bmatrix} \!\!
\end{equation}
where $c(i)$, slightly abusing notation, denotes the type of bus $i$, which also represents the type of subsystem $i$ as defined previously, depending on the context; 
$\bm{x}_i \!\!=\!\! [\theta_i, \omega_i, v_i ]\T$ with the entries being voltage angle, angular frequency and voltage magnitude of bus $i$, respectively; the notation $\Delta$ denotes the deviation from the equilibrium point; 
for each generator bus $i \in \mathcal{V}_1$, $\bm{\rho}_i \!\!=\!\! [K_{{\rm p}, i}, K_{{\rm q}, i}, \tau_{{\rm p}, i}, \tau_{{\rm q}, i}]\T$ with the entries being the frequency and voltage droop gains, and the time constants of low-pass filters for active and reactive power measurements of the inverter, respectively; 
for each load bus $i \!\in\!\! \mathcal{V}_2$, $\bm{\rho}_i \!=\!\! [S_{{\rm pf}, i}, S_{{\rm pv}, i}, S_{{\rm qf}, i}, S_{{\rm qv}, i} ]\T$ with the entries denoting the frequency and voltage sensitivities of the active (or reactive) load; $\bm{\rho}_{ij} = [R_{ij}, X_{ij}]\T$ with the entries denoting the resistance and reactance of line $(i,j)$; 
$\bm{E}_1$ is the $3\! \times \!3$ identity matrix; $\bm{E}_2$ is the $3\! \times \!3$ matrix with 1 in the top-left corner and 0 elsewhere; $\bm{A}_{c(i)}(\bm{\rho}_i)$ and $\bm{A}_{c(i)c(j)}(\bm{\rho}_{i}, \bm{\rho}_{ij})$ are matrix functions given as follows: 
\vspace{-3pt}
\begin{equation}
    \vspace{-3pt}
    \bm{A}_1\!=\!\! \begin{bmatrix}
        0 &\!\!\! \omega_{\rm b}                &\!\!\!\!\! \!0 \\
        0 &\!\!\! - \frac{1}{\tau_{{\rm p}, i}} &\!\!\!\!\! \!0 \\
        0 &\!\!\! 0                             &\!\!\!\!\! \!- \frac{1}{\tau_{{\rm q}, i}} 
    \end{bmatrix}, 
    \bm{A}_2 \!=\!\! \begin{bmatrix}
        0 &\!\! \omega_{\rm b}  &\!\! 0 \\
        0 &\!\! -S_{{\rm pf}, i} &\!\! -S_{{\rm pv}, i} \\
        0 &\!\! -S_{{\rm qf}, i} &\!\! -S_{{\rm qv}, i} 
    \end{bmatrix}
\end{equation}
\begin{equation}
    \vspace{-3pt}
    \bm{A}_{11} \!=\!\bm{A}_{12} \!\!=\! \begin{bmatrix}
        \! 0 & 0 & 0 & 0 & 0 & 0 \\
        \! \frac{B_{ij}  K_{{\rm p}, i}}{\tau_{{\rm p}, i}} \!\!\!&\!\!\! 0 \!\!\!&\!\!\!  \frac{G_{ij}  K_{{\rm p}, i}}{-\tau_{{\rm p}, i}} \!\!\!&\!\!\! 
        \frac{B_{ij}  K_{{\rm p}, i}}{- \tau_{{\rm p}, i}}  \!\!\!&\!\! 0 \!\!\!&\!\!\!  \frac{G_{ij}  K_{{\rm p}, i}}{\tau_{{\rm p}, i}}\\
        \! \frac{G_{ij}  K_{{\rm q}, i}}{\tau_{{\rm q}, i}} \!\!\!&\!\!\! 0 \!\!\!&\!\!\!  \frac{B_{ij}  K_{{\rm q}, i}}{\tau_{{\rm q}, i}} \!\!\!&\!\!\! 
        \frac{G_{ij}  K_{{\rm q}, i}}{-\tau_{{\rm q}, i}}  \!\!\!&\!\!  0 \!\!\!&\!\!\!  \frac{B_{ij}  K_{{\rm q}, i}}{-\tau_{{\rm q}, i}}  \\
    \end{bmatrix}
\end{equation}
\begin{equation}
    \vspace{-3pt}
    \bm{A}_{22}  = \bm{A}_{21} = \begin{bmatrix}
        0 & 0 & 0 & 0 & 0 & 0 \\
        B_{ij} \!\!\!&\!\!\! 0 \!\!\!&\!\!\!  -G_{ij}    \!&\!\!\!
        -B_{ij}  \!\!\!&\!\!\! 0 \!\!\!&\!\!\!  G_{ij}   \\
        G_{ij} \!\!\!&\!\!\! 0 \!\!\!&\!\!\!  B_{ij}    \!&\!\!\! 
        -G_{ij}  \!\!\!&\!\!\! 0 \!\!\!&\!\!\!  -B_{ij}     \\
    \end{bmatrix}
\end{equation}
with $G_{ij} \!=\!  {R_{ij}}/({R_{ij}^2 \!+\! X_{ij}^2})$, $B_{ij} \!=\! -{X_{ij}}/({R_{ij}^2 \!+\! X_{ij}^2})$, and $\omega_{\rm b}$ being the base frequency that is invariant for the class of type-consistent microgrids. 

Combining the individual small-signal model of all buses gives the small-signal model of the system, written as:  
\vspace{-3pt}
\begin{equation}\label{eq-8-1-3}
    \vspace{-3pt}
    \bm{E} \Delta \dot{\bm{x}} = \bm{A}(\bm{\rho}) \cdot \Delta \bm{x}
\end{equation}
where $\Delta \bm{x}$ is formed by concatenating $\Delta \bm{x}_i$ of all buses; 
$\bm{\rho}$ is formed by concatenating $\bm{\rho}_i$ of all buses and $\bm{\rho}_{ij}$ of all lines; 
the matrix $\bm{E}$ is formed with $\bm{E}_{c(i)}$ of all buses, and the matrix function $\bm{A}(\bm{\rho})$ is formed with $\bm{A}_{c(i)}$ of all buses and $\bm{A}_{c(i), c(j)}$ of all pairs of connected buses. 

The sufficient and necessary conditions for small-disturbance stability of the microgrid, i.e., asymptotic stability of system (\ref{eq-8-1-3}), can be expressed using generalized eigenvalues as follows \cite{4-rp-14}: 
\begin{theorem}
    The system (\ref{eq-8-1-3}) is asymptotically stable iff all generalized eigenvalues of the pair $(\bm{E}, \bm{A}(\bm{\rho}) )$ exhibit a negative real part, i.e, $\lambda_{\max}(\bm{E}, \bm{A}(\bm{\rho})) < 0$ with $\lambda_{\max}(\cdot)$ being the largest real part of the generalized eigenvalues of the pair. 
\end{theorem}


\subsection{Construction of the Neural DSC}

The existing DSCs for power systems analytically derived utilize different forms of local information of system dynamics. For instance, the DSCs in \cite{4-rp-11} and \cite{4-rp-12} are defined for each bus, in terms of the local dynamics of the bus (i.e., $\bm{A}_{c(i)}$) and coupling dynamics of its connected lines (i.e., $\bm{A}_{c(i)c(j)}$ for all $j \in \mathcal{A}_i$). 
This form of local system dynamics information is employed in the construction of the neural DSC. Alternative forms can also be utilized to construct neural DSCs in a similar fashion. 
Accordingly, the decomposition units for constructing the neural DSC are defined as buses and directed branches with terminal buses, yielding the following subsystem type set: 
\vspace{-3pt}
\begin{equation}
    \vspace{-3pt}
    \mathcal{C} = \mathcal{A} \cup \mathcal{F} = \{1,2, 1\text{-}1, 2\text{-}2, 1\text{-}2, 2\text{-}1 \}
\end{equation}
The sets of subsystems of each type are given as follows: 
\vspace{-3pt}
\begin{equation}
    \vspace{-3pt}
    \begin{aligned}
        & \mathcal{S}_1 \!\!=\!\! \mathcal{V}_1, \mathcal{S}_2 \!\!=\!\! \mathcal{V}_2, 
        \mathcal{S}_{1\text{-}1} \!\!=\!\! \{(i,j) \!\!\in\! \mathcal{V}_1 \!\!\times\!\! \mathcal{V}_1  | (i,j) \!~\text{or}~\! (j,i) \!\in\! \mathcal{E} \} \\
        & \mathcal{S}_{2\text{-}2} \!=\! \{(i,j) \!\in\! \mathcal{V}_2 \!\times\! \mathcal{V}_2  | (i,j) \!~\text{or}~\! (j,i) \!\in\! \mathcal{E}         \} \\
        & \mathcal{S}_{1\text{-}2} \!=\! \{(i,j) \!\in\! \mathcal{V}_1 \!\times\! \mathcal{V}_2  | (i,j) \!~\text{or}~\! (j,i) \!\in\! \mathcal{E}         \} \\
        & \mathcal{S}_{2\text{-}1} \!=\! \{(i,j) \!\in\! \mathcal{V}_2 \!\times\! \mathcal{V}_1  | (i,j) \!~\text{or}~\! (j,i) \!\in\!\mathcal{E}          \}
    \end{aligned}   
\end{equation}

\begin{figure}[h!]
    \vspace{-5pt}
	\centering 
    \includegraphics[scale=0.73]{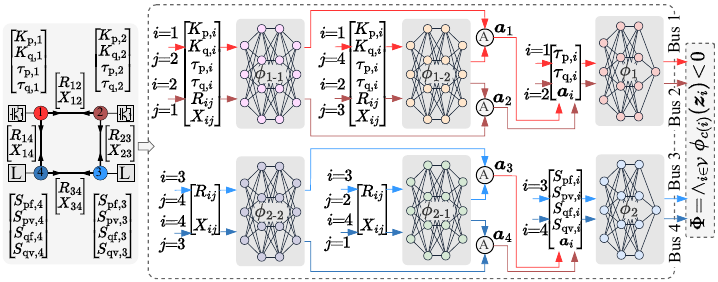}   
    \vspace{-2pt}
	\caption{Illustration of the neural decentralized stability condition.}
	\label{fig-8-1p2-6}
\end{figure}

After the system decomposition, a total of $N_{\rm nn}\!\!=\!\!|\mathcal{C}|\!\!=\!6$ NNs are introduced to construct the neural DSC, including $\phi_{1\text{-}1}(\bm{z}_i|\bm{\xi}_{1\text{-}1})$, $\phi_{2\text{-}2}(\bm{z}_i|\bm{\xi}_{2\text{-}2})$, $\phi_{1\text{-}2}(\bm{z}_i|\bm{\xi}_{1\text{-}2})$, and $\phi_{2\text{-}1}(\bm{z}_i|\bm{\xi}_{2\text{-}1})$, for the four types of subsystems that are directed branches with terminal buses; $\phi_{1}(\bm{z}_i|\bm{\xi}_1)$ and $\phi_{2}(\bm{z}_i|\bm{\xi}_2)$, for the two types of subsystems that are buses. 
They are all designed as FNNs. 
Fig. \ref{fig-8-1p2-6} illustrates the constructed neural DSC using a 4-bus microgrid where buses 1 and 2 are type-1 buses, and buses 3 and 4 are type-2 buses. 

First, to uniformly define the neural DSC for each bus in terms of its associated matrix $\bm{A}_{c(i)}$ and $\bm{A}_{c(i)c(j)}$ of all lines connecting it, the NNs $\phi_{1\text{-}1}$, $\phi_{2\text{-}2}$, $\phi_{1\text{-}2}$, and $\phi_{2\text{-}1}$ are designed to map the matrix function $\bm{A}_{c(i)c(j)}$ of each associated directed branch into the same representation space. In this way, the coupling dynamics of different types of directed branches for each bus can be handled uniformly.
Moreover, for subsystems of the same type, the associated matrix functions $\bm{A}_{c(i)}$ or $\bm{A}_{c(i)c(j)}$ share the same functional form but are evaluated at different input values. Thus, only the input to the matrix function is required for the corresponding NN. 
Specifically, we have 
\vspace{-3pt}
\begin{equation}
    \vspace{-3pt}
    \bm{o}_{ij} = \phi_{k}( \bm{z}_{ij} | \bm{\xi}_k) ~~ \forall k \in \mathcal{F}, (i, j) \in \mathcal{S}_k
\end{equation}
where the input $\bm{z}_{ij} = [ K_{{\rm p}, i}, K_{{\rm q}, i}, \tau_{{\rm p}, i}, \tau_{{\rm q}, i}, R_{ij}, X_{ij} ]\T$ for the NNs $\phi_{1\text{-}1}$ and $\phi_{1\text{-}2}$, and $\bm{z}_{ij} = [ R_{ij}, X_{ij} ]\T$ for the NNs $\phi_{2\text{-}1}$ and $\phi_{2\text{-}2}$; 
the output $\bm{o}_{ij} \in \mathbb{R}^{n_{\rm o}}$ is the $\eta_{\rm o}$-dimensional representation vector for subsystem $(i, j)$. 

Then, the rule-invariant operation aggregates the representation vectors of the directed branches connected to each bus as follows: 
\vspace{-3pt} 
\begin{equation}
    \vspace{-3pt}
    \bm{a}_i \!=\!  [\medmath{\sum_{j \in \mathcal{A}_i}} \! \bm{o}_{ij}\T, \frac{1}{|\mathcal{A}_i|}  \medmath{\sum_{j \in \mathcal{A}_i}} \!\bm{o}_{ij}\T,  \max\nolimits_{j \in \mathcal{A}_i} \bm{o}_{ij}\T  ]\T ~~ \forall i \!\in\! \mathcal{S}_1 \!\cup \mathcal{S}_2
\end{equation}
with $\bm{a}_i \in \mathbb{R}^{3 n_{\rm o}}$ being the aggregated vector for subsystem $i$. 

Furthermore, the NNs $\phi_{1}(\bm{z}_i|\bm{\xi}_1)$ and $\phi_{2}(\bm{z}_i|\bm{\xi}_2)$ map the input of $\bm{A}_{c(i)}$ and the aggregated vector $\bm{a}_i$ of each bus into a scalar to indicate satisfaction of the stability condition at the bus.   
Formally, for each bus, its stability condition candidate is represented as
\vspace{-3pt} 
\begin{equation}
    \vspace{-3pt}
    \phi_{k}( \bm{z}_{i} | \bm{\xi}_k )  < 0 ~~ \forall k \in \mathcal{A}, i \in \mathcal{S}_k 
\end{equation} 
where the input vector 
$\bm{z}_i = [\tau_{{\rm p}, i}, \tau_{{\rm q}, i}, \bm{a}_i\T]\T$ for $\phi_1$, and 
$\bm{z}_i = [S_{{\rm pf}, i}, S_{{\rm pv}, i}, S_{{\rm qf}, i}, S_{{\rm qv}, i}, \bm{a}_i\T]\T$ for $\phi_2$. 

Finally, the neural DSC is constructed as follows: 
\vspace{-3pt}
\begin{equation}
    \vspace{-3pt}
    \Phi(\bm{\rho} | \bm{\xi}) = \medmath{\bigwedge\nolimits_{i \in \mathcal{V}}}  \phi_{c(i)}( \bm{z}_{i} | \bm{\xi}_{c(i)} )  < 0
\end{equation}
where $\Phi = 1$ if $\phi_{c(i)} < 0$ holds for all buses, indicating that the small-disturbance stability of the microgrid is guaranteed.

\subsection{Training of the Neural DSC}

The generalizability of the neural DSC w.r.t. topological structure variations can be achieved similarly to the neural EF by training on a dataset covering a wide range of topological structures and validating on unseen ones. 
However, given the simper structure of the DSC, we are particularly interested in \textit{whether the neural DSC trained on a single system can generalize to other topological structures}. Thus, the training of the neural DSC is designed for this purpose. 
Considering a given topological structure, the region of interest for $\bm{\rho}$ in defining the DSC is denoted as $\mathbb{P}$. This region is further divided into two parts: the instability region $\mathbb{P}_{1}$, where system (\ref{eq-8-1-3}) is not asymptotically stable; and the stability region $\mathbb{P}_{2}$, where system (\ref{eq-8-1-3}) is asymptotically stable. 
Then the construction of the neural DSC amounts to finding the value of $\bm{\xi}$, such that for any $\bm{\rho} \in \mathbb{P}$: 
\vspace{-3pt}
\begin{equation}\label{eq-8-1-5}
    \vspace{-3pt}
    \Phi(\bm{\rho} | \bm{\xi}) = 1 \!\Rightarrow\! \lambda_{\max}(\bm{E}, \bm{A}(\bm{\rho})) \!<\! 0
\end{equation} 
and the neural DSC is minimally conservative.

\subsubsection{Training set} 
The training set is formed by $N_1 + N_2$ samples of $\bm{\rho}$, with the first $N_1$ samples in region $\mathbb{P}_1$ and the remaining $N_2$ samples in $\mathbb{P}_2$. This can be carried out by randomly sampling from the region $\mathbb{P}$ and computing $\lambda_{\max}(\bm{E}, \bm{A}(\bm{\rho}))$ for each sample.

\subsubsection{Model Training}

According to the proposition (\ref{eq-8-1-5}), we can yield the following two corollaries: 
\begin{itemize}

    \item The probability that the left-hand side of (\ref{eq-8-1-5}) holds for any randomly chosen $\bm{\rho} \in \mathbb{P}_1$, denoted as $P_1$, is 0. This is the equivalent contrapositive of proposition (\ref{eq-8-1-5}) and should therefore be guaranteed strictly.

    \item The probability that the left-hand side of (\ref{eq-8-1-5}) does not hold for any $\bm{\rho} \!\in\! \mathbb{P}_2$, denoted as $P_2$, is indeterminate. This probability quantifies the DSC' conservativeness, with smaller values indicating lower conservativeness overall.

\end{itemize}

Accordingly, the value of $\bm{\xi}$ with which the proposition (\ref{eq-8-1-5}) holds and the DSC are minimally conservative for all training samples, is given by the following optimization problem: 
\vspace{-3pt}
\begin{equation}\label{eq-8-1-7}
    \vspace{-3pt}
    \begin{aligned}
        \min\nolimits_{\bm{\xi}} ~ & \hat{P}_2 =   \tfrac{1}{ N_2 } \medmath{\sum\nolimits_{ l = N_1+1 }^{N_1 + N_2}}  \max_{i \in \mathcal{V}}   H\big(  \phi_{c(i)}( \bm{z}_{i}^l | \bm{\xi}_{c(i)} ) \big)   \\ 
        \text{s.t.} ~  & \hat{P}_1 \!=\! \tfrac{1}{ N_1 } \! \medmath{\sum\nolimits_{ l = 1 }^{N_1}}  \Big[ 1 \!-\! \max_{i \in \mathcal{V}} H \big( \phi_{c(i)}( \bm{z}_{i}^l | \bm{\xi}_{c(i)} ) \big)  \Big] \!=\! 0
    \end{aligned}
\end{equation}
where $H(\cdot)$ is the unit step function that equals to $0$ for negative arguments and 1 otherwise; 
$\bm{z}_{i}^{l}$ denotes the value of $\bm{z}_{i}$ of the $l$-th training sample. 
The terms $\hat{P}_1$ and $\hat{P}_2$ are the empirical values of the probability $P_1$ and $P_2$, respectively.

Based on problem (\ref{eq-8-1-7}), the loss function $L(\bm{\xi})$ for training all NNs of the neural DSC is defined as the same form as Eq. (\ref{eq-8-1p2-1}) in the proposed NN paradigm, with each component defined as: 
\vspace{-3pt}
\begin{equation}\label{eq-8-1p2-a1}
    \vspace{-3pt}
    L_1 = - \tfrac{1}{ N_1 } \medmath{\sum\nolimits_{ l = 1 }^{N_1}} \log \Big[ \max_{i \in \mathcal{V}} \sigma \big( \phi_{c(i)}( \bm{z}_{i}^l | \bm{\xi}_{c(i)} ) \big)  \Big]
\end{equation}
\vspace{-3pt}
\begin{equation}\label{eq-8-1p2-a2}
    L_2 = - \tfrac{1}{ N_2 }  \medmath{\sum\nolimits_{ l = N_1+1 }^{N_1 + N_2}} \log \Big[ 1 - \max_{i \in \mathcal{V}} \sigma \big( \phi_{c(i)}( \bm{z}_{i}^l | \bm{\xi}_{c(i)} ) \big)  \Big]
\end{equation}
\vspace{-6pt}
\begin{equation}\label{eq-8-1p2-a3}
    \vspace{-3pt}
    \!\!\! L_{\rm aux}  \!\!=\!\! \frac{1}{N_1 \!\!+\!\! N_2} \!\!\! \medmath{\sum\limits_{ l = 1 }^{N_1 \!+\! N_2}}  \!\!\! \big[\!  \max\limits_{i \in \mathcal{V}}  \phi_{c(i)}(\! \bm{z}_{i}^l | \bm{\xi}_{c(i)} \!) \!-\!\! \lambda_{\max}(\bm{E},\! \bm{A}(\bm{\rho}^l))  \big]^2  \!\!\!
\end{equation}
with $\sigma(\cdot)$ being the sigmoid function and $\bm{\rho}^l$ being the value of $\bm{\rho}$ of the $l$-th training sample. 
The terms $L_1$ and $L_2$, devised with inspiration from the cross-entropy loss, each serve as proxies for $\hat{P_1}$ and $\hat{P_2}$. 
The term $L_{\rm aux}$ represents the average distance between the largest value of $\phi_{c(i)}( \cdot )$ among all buses (roughly correlating with the system's stability level), and $\lambda_{\max}(\cdot)$, over all samples.

\subsubsection{Verification}
The verification of the neural DSC with the trained NNs is performed by estimating the probability that proposition (\ref{eq-8-1-5}) holds for any $\bm{\rho} \in \mathbb{P}$, denoted as $P_3$. 
Specifically, generate $N_{\rm v}$ samples of $\bm{\rho}$ where the left-hand side of (\ref{eq-8-1-5}) holds to form a verification set. 
Let $N_{\rm c}$ denote the number of counterexamples, i.e., validation samples where $\lambda_{\max}(\bm{E}, \bm{A}(\bm{\rho} )) \!\geq\! 0$. 
Then the empirical values of $P_3$ on the verification set is expressed as $\hat{P}_3^{\rm v} \!\!=\!\! (N_{\rm v} \!-\! N_{\rm c})/{N_{\rm v}}$. The verification is passed if $\hat{P}_3^{\rm v} = 1$.

\subsubsection{Validation}

Two validation sets, name A and B is first constructed. Set A is analogous to the verification set, but contains  $N_{\rm A}$ samples, which is significantly larger than the verification set size $N_{\rm v}$. 
The empirical value of $P_3$ on validation set A, denoted as $\hat{P}_{3}^{\rm q}$, is computed analogously to $\hat{P}_3^{\rm v}$. 
Validation set B contains all counterexamples found in validation set A. 
An empirical value of $P_1$, denoted as $\hat{P}_1^{\rm q}$, can be easily obtained on this validation set. 
It is noted that in each training round, validation set A needs to be regenerated since the NNs are changed. 
Validation set B combines all counterexamples found in validation set A from both the current and all previous training rounds. 
The validation is passed if $\hat{P}_1^{\rm q} = 0$ and $\hat{P}_3^{\rm q} = 1$, indicating that the obtained neural DSC is almost certainly valid. 

\subsubsection{Sample augmentation}
Following the failed verification, sample augmentation adds the identified counterexamples of $\bm{\rho}$ together with $N_{\rm p}$ samples around them to the training set. 
Following the failed validation, the verification set is regenerated to identify counterexamples, which together with $N_{\rm p}$ samples around them, are added to the training set.

\vspace{-1pt}
\section{Numerical Results}

This section demonstrates the applicability and effectiveness of the proposed NN paradigm through the simulation results for stability analysis of bulk power grids and microgrid systems. 

\vspace{-3pt}
\subsection{Bulk Power Grid Stability Analysis -- Results of the Neural Energy Function}

The implementation of the proposed NN paradigm for finding EFs is numerically tested using the parameter setting given in Table \ref{tab-5-8-1p2-1}. 
All parameter values for $\mathbb{X}_s$ and $\mathbb{P}_s$ are expressed in per unit, except for $\theta_i$ and $\delta_i$, which are in radians.

\begin{table}[h!]
    \vspace{-5pt}
    \centering
    \caption{Parameter setting for the neural EF}
    \setlength{\tabcolsep}{5pt} 
    \renewcommand{\arraystretch}{1.2}

    \small{
    \begin{tabular*}{\hsize}{c|l}  \toprule  \addlinespace[-0.5pt]
     $\mathbb{X}_s$ &  $\theta_i, \delta_i \!\in\! [-2\pi, 2\pi]$, $\omega_i \!\in\! [-0.2, 0.2]$ \\ \hline
     $\mathbb{P}_s$ &   \makecell[l]{ \vspace{-8pt} \\ $p_{{\rm g}, i}, p_{{\rm d}, i}\!\in\! [0.5, 1.5]$, $\!\sum_{i \in \mathcal{V}_3} \! p_{{\rm d}, i} \!-\! \sum_{i \in \mathcal{V}_2} \! p_{{\rm g}, i} \!\in\! [0.5, 1.5]$, \\ $b_i, b_{ij}\!\in\! [\frac{1}{0.3}, \frac{1}{0.1}]$, $m_i \!\in\! [2, 10]$,  $d_{{\rm g}, i}\!\in\! [4, 20]$, $d_{{\rm d}, i} \!\in\! [4, 6]$ \\  \vspace{-9pt} } \\ \hline
     FNNs & \makecell[l]{ \vspace{-8pt} \\ SiLU activation functions; 3 hidden layers with widths \\ of 8, 8, and 16 times the input dimension, respectively. \\ \vspace{-9pt}}  \\ \hline
     Train &  \makecell[l]{\vspace{-8pt} \\ $\varphi \!=\!\! 10^{-2}\!$, $\varphi' \!\!=\! 10^{-5}\!$, $\varepsilon \!=\!\! 10^{-3}$, $\eta_1\!=\!\!10$, $\eta_2\!=\!20$, $\eta_3\!=\!\!50$, \\ $\eta_4^s\!=\!\!2 \!\times\! 10^4$, $\eta_5\!=\!\!10$, $|\mathcal{X}_{\rm b}|\!=\!\!100$, $\eta_6\!=\!\!50$, $\eta_7\!=\!\!100$, \\ $\eta_8\!=\!10^8$, $\eta_9\!=\!\!99$, $\eta_{10}\!=\!20$, batch size $\!=\!$ 51200, \\
     optimizer: Adam with step decay of learning rate \\ \vspace{-9pt} }   \\[-0.55mm] \bottomrule
    \end{tabular*} 
    }
    \label{tab-5-8-1p2-1}  
    \vspace{-5pt}
\end{table}

\subsubsection{Training process} 
\begin{figure}[h!]
	\centering 
    \includegraphics[scale=1.0]{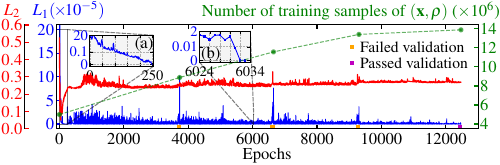}   
	\caption{Curves of $L_1$, $L_2$, and training sample size for $(\bm{x}, \bm{\rho})$  during training.}
	\label{fig-8-2-r3}
\end{figure}

Fig. \ref{fig-8-2-r3} shows the curves of $L_1$, $L_2$, and the training sample size for $(\bm{x}, \bm{\rho})$. 
The initial training round reduces $L_1$ to 0 after approximately 250 epochs, as shown in Fig. \ref{fig-8-2-r3}-(a). 
Differently, each subsequent training round achieves the same minimization of $L_1$ with notably fewer epochs—such as 10 epochs, as shown in Fig. \ref{fig-8-2-r3}-(b). 
The training set is augmented three time following failed validation, and the training sample size for $(\bm{x}, \bm{\rho})$ is increased from $5 \!\times\! 10^6$ to nearly $14 \!\times\! 10^6$. 
After about 12,000 epochs, the passed validation concludes the entire training, with a reduction of $L_2$ approximately from 0.5 to 0.3. 

\subsubsection{Generalizability to post-fault system states} 
\begin{figure}[h!]
	\centering 
    \includegraphics[scale=0.99]{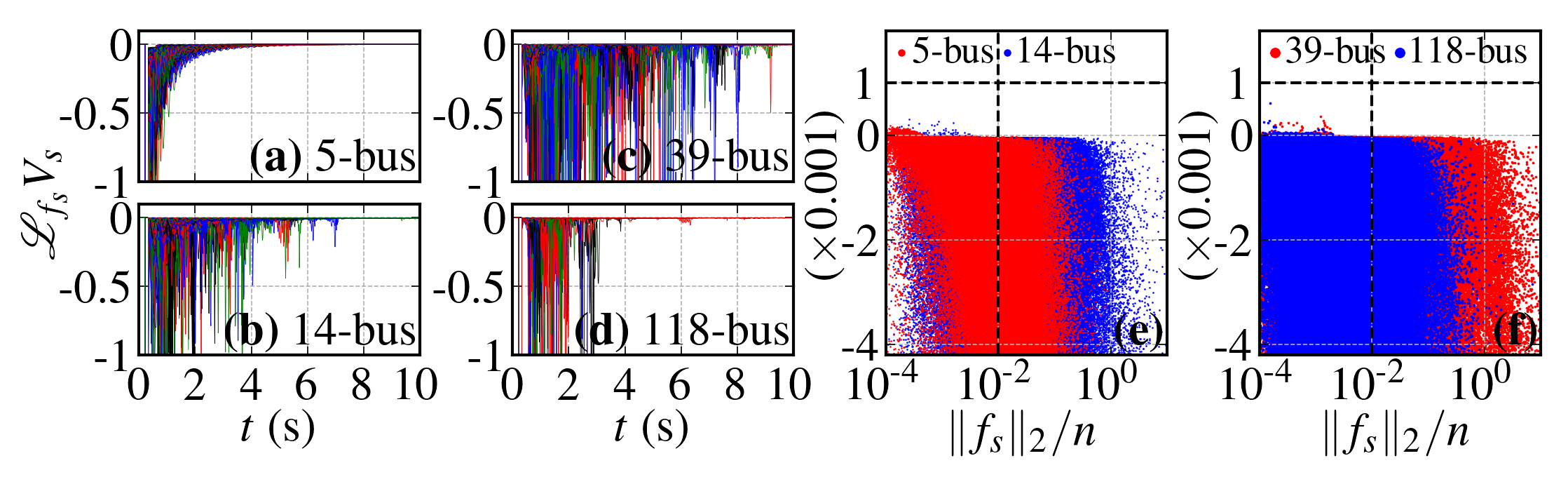}   
	\caption{(a)-(d) Time-domain curves of $\mathcal{L}_{f_s} V_s$ for the four test systems under various fault scenarios; (e)-(f) scatter plots of $\Vert f_s \Vert_2 / n$-$\mathcal{L}_{f_s}\! V_s$ after the faults. }
	\label{fig-8-2-r2}
\end{figure}

The generalizability of the obtained EF is further demonstrated via post-fault system state trajectories from 4 systems which are not used in either the training or validation set in the iterative training process. 
These systems are the 5-, 14-, 39-, and 118-bus systems, with the same topological structure as the equal-sized IEEE test systems. 
For each system, 1000 different scenarios of $\bm{\rho}$ values taken from $\mathbb{P}_s$ and 3-phase short-circuit faults (occurring at $t\!=\!0.1$ s and clearing after 0.1 s)
are used, in each of which the system is stable. 
Fig. \ref{fig-8-2-r2}-(a) to (d) show the time-domain curves of $\mathcal{L}_{f_s} V_s$ under these scenarios. 
Intuitively, the value of $\mathcal{L}_{f_s} V_s$ increases but remains non-positive after the faults, and converges to 0 as time progresses. 
This is consistent with the expected change of EF values after faults. 
Precisely, Fig. \ref{fig-8-2-r2}-(e) and (f) give the scatter plots of $\Vert f_s \Vert_2/n$-$\mathcal{L}_{f_s} V_s$ for all post-fault state points. 
It is seen that for these state points, $\mathcal{L}_{f_s} V_s \!\!<\!\! 0$ if $\Vert f_s \Vert_2/n \!\geq\! 0.01 \!\!=\!\! \varphi$, and $\mathcal{L}_{f_s} V_s \!\!<\!\! 0.001 \!\!=\!\! \varepsilon$ otherwise. Thus the conditions for EFs are satisfied with the pre-set tolerance. 

\begin{figure}[h!]
	\centering 
    \includegraphics[scale=1]{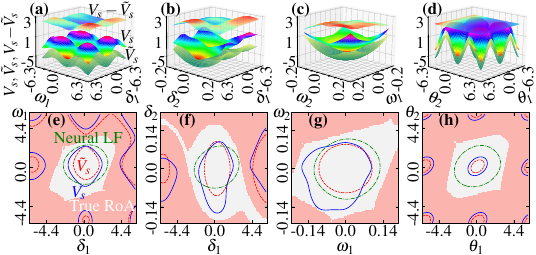}   
	\caption{(a)-(d) Surface plots of the neural EF at different cross-sections; (e)-(h) comparison of the estimated RoAs at different cross-sections. }
	\label{fig-8-2-r1}
    \vspace{-9pt}
\end{figure}

\subsubsection{Comparison of RoAs} 
Taking the 5-bus system as an example, Fig. \ref{fig-8-2-r1}-(a) to (d) visualize the obtained neural EF $V_s$, the analytical EF $\tilde{V}_s$, and the NN component of $V_s$, i.e., $V_s \!-\! \tilde{V}_s$, at different cross-sections. 
It can be seen that at each cross-section, the overall shape of $V_s$ is similar to $\tilde{V}_s$, but their specific values are entirely different due to the NN components. 
Moreover, Fig. \ref{fig-8-2-r1}-(e) to (h) compare the estimated RoAs obtained using the neural LF in \cite{4-1377}, $\tilde{V}_s$, and $V_s$, corresponding to the white regions enclosed by the green, red, and blue curves, respectively. 
It is observed that the neural EF $V_s$ yields a larger RoA than the analytical EF $\tilde{V}_s$ does. 
At the cross-sections of $\delta_1$-$\delta_2$ and $\omega_1$-$\omega_2$, the size of the RoA estimated using $V_s$ is close to that obtained using the neural LF, which although is specifically constructed for this 5-bus system and not scalable. 
This observation indicates that the neural EF yields less conservative stability results than the analytical EF, while remaining scalable.

\subsection{Microgrid System Stability Analysis -- Results of the Neural Decentralized Stability Condition}

\begin{figure}[t!]
	\centering 
    \includegraphics[scale=0.83]{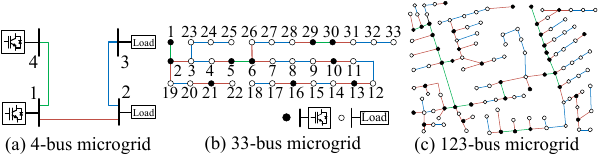}  
    \vspace{-4pt}
	\caption{Diagrams of the three type-consistent microgrids.}
	\label{fig-8-1-5}
\end{figure}

The implementation of the proposed NN paradigm for constructing DSCs is numerically tested using the three microgrid systems shown in Fig. \ref{fig-8-1-5}. These microgrids exhibit different topological structures and sizes, but all belong to the class of type-consistent microgrids introduced in Section \ref{sec-8-1p2-4}. The neural DSC is constructed for each microgrid separately with the parameter setting given in Table \ref{tab-5-8-1p2-2}.

\subsubsection{Training process}
\begin{figure}[t!]
	\centering 
    \includegraphics[scale=1.17]{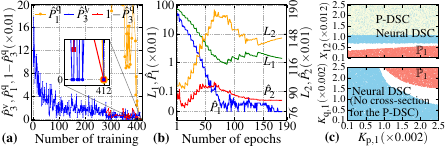}  
    \vspace{-4pt}
	\caption{(a)-(b) Curves of the associated terms during the training; (c) cross-sections of the stability regions formed by different DSCs for the 4-bus system.}
	\label{fig-8-1-6}
\end{figure}

Fig. \ref{fig-8-1-6}-(a) and \ref{fig-8-1-6}-(b), taking the 123-bus microgrid for instance, respectively show the curves of $\hat{P}_3^{\rm v}$, $\hat{P}_1^{\rm q}$, and $\hat{P}_3^{\rm q}$ during the entire iterative training process, and the curves of $L_1$, $L_2$, $\hat{P}_1$, and $\hat{P}_2$ for the last training round. 
From Fig. \ref{fig-8-1-6}-(b), it is seen that during the last training round, within around the first 100 epochs, $L_1$ decreases from 1 to about 0.01, while $\hat{P}_1$ drops from 1 to nearly 0. Then, $L_1$ exhibits a slight increase, while $\hat{P}_1$ fluctuates around 0 and finally converges to 0. Meanwhile, $L_2$ and $\hat{P}_2$ increase during the early training stage, followed by a decreasing trend, with $\hat{P}_2$ eventually decreasing to about 0.85. 
These trajectories demonstrate that the problem (\ref{eq-8-1-7}) is effectively solved by training the NNs using the devised loss function. 
Also, as shown in Fig. \ref{fig-8-1-6}-(a), $\hat{P}_3^{\rm v}$ decreases overall as training rounds progress and first attains 0 within fewer than 300 rounds, marking the initial passing of verification. Subsequently, $\hat{P}_1^{\rm q}$ from the validation decreases overall and reaches 0 at the 412th training round, where $\hat{P}_3^{\rm q}$ attains 1. This indicates that the validation is passed and the iterative training scheme produces the valid DSC for the microgrid.

\begin{table}[t!]
    \vspace{-5pt}
    \centering
    \caption{Parameter setting for the neural DSC}
    \vspace{-4pt}
    \setlength{\tabcolsep}{5pt}  
    \renewcommand{\arraystretch}{1.3}
    \small{
    \begin{tabular*}{\hsize}{c|l} \toprule \addlinespace[-0.8pt]
     $\mathbb{P}$ &   \makecell[l]{ \vspace{-8pt} \\  $K_{{\rm p}, i}, K_{{\rm q}, i} \!\!\in\!\! [0.0002, 0.005]$, $\tau_{{\rm p}, i}, \tau_{{\rm q}, i} \!\!\in\!\! [0.006, 0.15]$,  \\
     $R_{ij} \!\!\in\!\! [0.0001, 0.0025]$, $X_{ij} \!\!\in\!\! [0.0002, 0.005]$, \\
     $S_{{\rm pv}, i}, S_{{\rm qf}, i} \!\!\in\!\! [0.0002, 0.005]$, $S_{{\rm pf}, i}, S_{{\rm qv}, i} \!\!\in\!\! [0.001, 0.025]$
     \\  \vspace{-9pt} } \\ \hline
     FNNs & \makecell[l]{ \vspace{-8pt} \\ ReLU activation functions; layer configuration: \\  (6,30,30,6) for $\phi_{1\text{-}1}\!$ and $\!\phi_{1\text{-}2}$, (2,10,10,2) for $\phi_{2\text{-}1}\!$ and $\!\phi_{2\text{-}2}$, \\ (20,100,100,100,1) for $\phi_{1}$,  (10,50,50,50,1) for $\phi_{2}$ \\ \vspace{-9pt}}  \\ \hline
     Train &  \makecell[l]{\vspace{-8pt} \\  $N_1 \!\!+\!\! N_2 \!\!=\!\! N_{\rm v} \!\!=\!\! 2\!\!\times\!\!{10}^4\!$, $\!N_{\rm A} \!\!=\! 4\!\!\times\!\!{10}^4\!$, $\!N_{\rm p} \!\!=\!\! 5$, $\!$batch size $\!\!=\!\!$ 2560, \\
     optimizer: Adam with step decay of learning rate  \\ \vspace{-9pt} }   \\[-0.55mm]  \bottomrule
    \end{tabular*} 
    }
    \label{tab-5-8-1p2-2}  
\end{table}

\begin{table}[t!]
    \centering
 
    \caption{Performance of the neural DSCs and the P-DSC }
    \vspace{-4pt}
    \setlength{\tabcolsep}{0.99pt} 
    \renewcommand{\arraystretch}{1.3}

    \setlength{\aboverulesep}{0.2pt}
    \setlength{\belowrulesep}{0.2pt}
    \setlength{\extrarowheight}{-.0ex}
 
    \footnotesize{
    \begin{tabular*}{\hsize}{@{}p{1.06cm}ccccccccc}\toprule
    \multirow{2}{*}{ DSCs  } & \multicolumn{3}{c}{$P_1$} & \multicolumn{3}{c}{$P_3$} & \multicolumn{3}{c}{$P_2$}  
    \\ \cmidrule(lr){2-4} \cmidrule(lr){5-7} \cmidrule(lr){8-10} 
                        & 4-bus       & 33-bus     & 123-bus    &  4-bus    & 33-bus   & 123-bus  & 4-bus & 33-bus & 123-bus      \\ \midrule[0.7pt]
    4-bus               &  \tblue0.00\% & 40.4\%       & 67.4\%       &  \tred100\% & 63.1\%     & 6.98\%     &  95.5\% & ---    & ---   \\
    33-bus              &  17.2\%       & \tblue0.00\% & 31.7\%       &  97.3\%     & \tred100\% & 23.6\%     &  ---  & 46.3\%    & ---   \\
    123-bus             &  \tblue0.00\% & \tblue0.00\% & \tblue0.00\% &  \tred100\% & \tred100\% & \tred100\% &  91.6\% & 38.2\%    & 15.9\%   \\
    P-DSC      &  \tblue0.00\% & \tblue0.00\% & \tblue0.00\% &  \tred100\% & \tred100\% & \tred100\% &  82.1\% & 25.6\%    & 6.55\%   \\ 
    123-bus$^*$         &  \tblue 2/2 & \tblue5/5 & \tblue40/42 &  \tblue 2/2 & \tblue5/5 & \tblue40/42 &  93.2\% & 38.5\%     & 16.1\%   \\ \bottomrule
    \multicolumn{10}{@{}p{1\columnwidth}@{}}{
        \footnotesize{\textit{Note}: The last row shows the proportion of $P_1 = 0$ for $P_1$, the proportion of $P_3 = 100\%$ for $P_3$, and the mean value for $P_2$, among all cases when the associated microgrid remove one bus without disconnecting the network. 
      }}
    \end{tabular*} 
    }
    \label{tab-5-8-1}  
    \vspace{-11pt}
\end{table}

\subsubsection{Generalizability} 
Table \ref{tab-5-8-1} compares the performance of both the obtained neural DSCs and the passivity-based DSC derived analytically in \cite{4-rp-12} (hereafter referred to as P-DSC). This evaluation is based on the empirical values of ${P}_1$, estimated on the final validation set B from the training for the associated microgrid, as well as $P_2$ and $P_3$, both computed using a 40000-sized sample set generated similarly to the training set. The neural DSCs established for each microgrid are evaluated on all the three microgrids. 
In Table \ref{tab-5-8-1}, the empirical values of $P_1$ and $P_2$ are 0 and 1, respectively, for the neural DSC obtained and evaluated using the same test system. This again confirms the validity of each neural DSC w.r.t. the topological structure used to train it, namely, the generalizability of the neural DSC to unseen values of system parameters. 

More importantly, the neural DSC obtained using the 123-bus microgrid is also valid w.r.t. the other two microgrids, given the unchanged empirical values of $P_1$ and $P_2$ evaluated on the other two microgrids. Also, the validity of the neural DSC obtained using the 123-bus microgrid still holds in nearly all cases when the network removes one bus, according to the results given in the last row of Table \ref{tab-5-8-1}. 
In contrast, the neural DSC obtained using the 4-bus microgrid or the 33-bus microgrid lacks such validity when applied to the other microgrids or to networks resulting from the removal of a single bus. 
Therefore, it can be concluded that by increasing the microgrid size used for training, the obtained neural DSC can become generalizable w.r.t. variations in both topological structures and system parameters.

\subsubsection{Conservativeness}
According to Table \ref{tab-5-8-1}, for each microgrid, the empirical values of $P_2$ corresponding to the neural DSCs are all larger than that corresponding to the P-DSC. This indicates that the obtained neural DSCs are all less conservative than the P-DSC. 
Intuitively, Fig. \ref{fig-8-1-6}-(c) displays two cross-sections of different stability regions for the 4-bus test system, where the cross-section dimensions are given by $K_{{\rm p}, i}$ and $K_{{\rm q}, i}$ for bus 1 in one case, and by $K_{{\rm p}, i}$ for bus 1 and $X_{ij}$ for line (1,2) in the other, with all other parameters fixed at ten times their respective minimal values. 
It can be observed that, in the first case, the P-DSC fails to yield stability results over the entire actual stability region, whereas the neural DSC yields a stability region (blue area) that is a subset of the true region. 
In the second case, both the neural DSC and the P-DSC yield valid stability regions, with the former covering a larger area. 
This observation demonstrates the lower conservativeness of the neural DSC compared to the analytically derived one.

\section{Conclusion}

This paper proposes a new NN paradigm to enable scalable and generalizable stability analysis of power systems, with the capability to reduce the conservativeness of stability results and to accommodate complex system dynamics. 
Two implementations of the NN paradigm respectively on large- and small-disturbance stability analysis demonstrate its effectiveness. 
The constructed neural EF is scalable and also exhibits generalizability to the considered class of type-consistent systems. 
It generates less conservative RoAs compared to those obtained using the analytical EF. 
Similarly, the constructed neural DSC is inherently scalable and also becomes generalizable to variations in both system parameters and topological structures when trained on a large-scale microgrid. The neural DSC yields notably larger stability regions than the analytical passivity-based DSC. 
Future work will focus on enhancing the sample-augmented iterative training scheme via meta-learning, exploring more paradigm implementations, and identifying previously undiscovered stability descriptor forms.

\ifCLASSOPTIONcaptionsoff
  \newpage
\fi

\bibliographystyle{IEEEtran}
\bibliography{./4.bib}

\end{document}